\PassOptionsToPackage{unicode}{hyperref}
\PassOptionsToPackage{hyphens}{url}
\PassOptionsToPackage{dvipsnames,svgnames,x11names}{xcolor}
\documentclass[
]{article}

\usepackage{amsmath,amssymb}
\usepackage{iftex}
\ifPDFTeX
  \usepackage[T1]{fontenc}
  \usepackage[utf8]{inputenc}
  \usepackage{textcomp} 
\else 
  \usepackage{unicode-math}
  \defaultfontfeatures{Scale=MatchLowercase}
  \defaultfontfeatures[\rmfamily]{Ligatures=TeX,Scale=1}
\fi
\usepackage{lmodern}
\ifPDFTeX\else  
\fi
\IfFileExists{upquote.sty}{\usepackage{upquote}}{}
\IfFileExists{microtype.sty}{
  \usepackage[]{microtype}
  \UseMicrotypeSet[protrusion]{basicmath} 
}{}
\makeatletter
\@ifundefined{KOMAClassName}{
  \IfFileExists{parskip.sty}{%
    \usepackage{parskip}
  }{
    \setlength{\parindent}{0pt}
    \setlength{\parskip}{6pt plus 2pt minus 1pt}}
}{
  \KOMAoptions{parskip=half}}
\makeatother
\usepackage{xcolor}
\ifx\paragraph\undefined\else
  \let\oldparagraph\paragraph
  \renewcommand{\paragraph}[1]{\oldparagraph{#1}\mbox{}}
\fi
\ifx\subparagraph\undefined\else
  \let\oldsubparagraph\subparagraph
  \renewcommand{\subparagraph}[1]{\oldsubparagraph{#1}\mbox{}}
\fi

\usepackage{longtable,booktabs,array}
\usepackage{calc} 
\usepackage{etoolbox}
\makeatletter
\patchcmd\longtable{\par}{\if@noskipsec\mbox{}\fi\par}{}{}
\makeatother
\IfFileExists{footnotehyper.sty}{\usepackage{footnotehyper}}{\usepackage{footnote}}
\makesavenoteenv{longtable}
\usepackage{graphicx}
\makeatletter
\def\maxwidth{\ifdim\Gin@nat@width>\linewidth\linewidth\else\Gin@nat@width\fi}
\def\maxheight{\ifdim\Gin@nat@height>\textheight\textheight\else\Gin@nat@height\fi}
\makeatother
\setkeys{Gin}{width=\maxwidth,height=\maxheight,keepaspectratio}
\makeatletter
\def\fps@figure{htbp}
\makeatother
\NewDocumentCommand\citeproctext{}{}

\makeatletter
 \let\@cite@ofmt\@firstofone
 \def\@biblabel#1{}
 \def\@cite#1#2{{#1\if@tempswa , #2\fi}}
\makeatother
\newlength{\cslhangindent}
\newlength{\csllabelwidth}
\newenvironment{CSLReferences}[2] 
 {\begin{list}{}{%
  \setlength{\itemindent}{0pt}
  \setlength{\leftmargin}{0pt}
  \setlength{\parsep}{0pt}
  \ifodd #1
   \setlength{\leftmargin}{\cslhangindent}
   \setlength{\itemindent}{-1\cslhangindent}
  \fi
  \setlength{\itemsep}{#2\baselineskip}}}
 {\end{list}}
\usepackage{calc}

\usepackage{arxiv}
\usepackage{orcidlink}
\usepackage{amsmath}
\usepackage[T1]{fontenc}
\usepackage{lmodern}
\makeatletter
\@ifpackageloaded{caption}{}{\usepackage{caption}}
\AtBeginDocument{%
\ifdefined\contentsname
  \renewcommand*\contentsname{Table of contents}
\else
  \newcommand\contentsname{Table of contents}
\fi
\ifdefined\listfigurename
  \renewcommand*\listfigurename{List of Figures}
\else
  \newcommand\listfigurename{List of Figures}
\fi
\ifdefined\listtablename
  \renewcommand*\listtablename{List of Tables}
\else
  \newcommand\listtablename{List of Tables}
\fi
\ifdefined\figurename
  \renewcommand*\figurename{Figure}
\else
  \newcommand\figurename{Figure}
\fi
\ifdefined\tablename
  \renewcommand*\tablename{Table}
\else
  \newcommand\tablename{Table}
\fi
}
\@ifpackageloaded{float}{}{\usepackage{float}}
\floatstyle{ruled}
\@ifundefined{c@chapter}{\newfloat{codelisting}{h}{lop}}{\newfloat{codelisting}{h}{lop}[chapter]}
\floatname{codelisting}{Listing}

\makeatother
\makeatletter
\@ifpackageloaded{caption}{}{\usepackage{caption}}
\@ifpackageloaded{subcaption}{}{\usepackage{subcaption}}
\makeatother
\ifLuaTeX
  \usepackage{selnolig}  
\fi
\usepackage{bookmark}

\IfFileExists{xurl.sty}{\usepackage{xurl}}{} 
\hypersetup{
  pdftitle={When Sensitivity Bias Varies Across Subgroups: The Impact of Non-uniform Polarity in List Experiments},
  pdfauthor={Sophia Hatz; David Randahl},
  pdfkeywords={List experiments, Sensitivity bias, Social desirability
bias, Simulation},
  colorlinks=true,
  linkcolor={blue},
  filecolor={Maroon},
  citecolor={Blue},
  urlcolor={Blue},
  pdfcreator={LaTeX via pandoc}}

\usepackage{setspace}
\newcommand{\runninghead}{A Preprint }
\renewcommand{\runninghead}{Working paper }
\title{When Sensitivity Bias Varies Across Subgroups: The Impact of
Non-uniform Polarity in List Experiments\thanks{The research was funded
by the European Research Council, project H2020-ERC-2015-AdG 694640
(ViEWS) and Riksbankens Jubileumsfond, grant M21-0002 (Societies at
Risk)}}
\def\asep{\\\\\\ } 
\author{\textbf{Sophia Hatz}\\Department of Peace and Conflict
Research\\Uppsala
University\\\\\href{mailto:sophia.hatz@pcr.uu.se}{sophia.hatz@pcr.uu.se}\asep\textbf{David
Randahl}~\orcidlink{0000-0003-1069-6067}\\Department of Peace and
Conflict Research\\Uppsala
University\\\\\href{mailto:david.randahl@pcr.uu.se}{david.randahl@pcr.uu.se}}
\date{}
\begin{document}
\maketitle
\begin{abstract}
Survey researchers face the problem of sensitivity bias: since people
are reluctant to reveal socially undesirable or otherwise risky traits,
aggregate estimates of these traits will be biased. List experiments
offer a solution by conferring respondents greater privacy. However,
little is know about how list experiments fare when sensitivity bias
varies across respondent subgroups. For example, a trait that is
socially undesirable to one group may socially desirable in a second
group, leading sensitivity bias to be negative in the first group, while
it is positive in the second. Or a trait may be not sensitive in one
group, leading sensitivity bias to be zero in one group and non-zero in
another. We use Monte Carlo simulations to explore what happens when the
polarity (sign) of sensitivity bias is non-uniform. We find that a
general diagnostic test yields false positives and that commonly used
estimators return biased estimates of the prevalence of the sensitive
trait, coefficients of covariates, and sensitivity bias itself. The bias
is worse when polarity runs in opposite directions across subgroups, and
as the difference in subgroup sizes increases. Significantly,
non-uniform polarity could explain why some list experiments appear to
`fail'. By defining and systematically investigating the problem of
non-uniform polarity, we hope to save some studies from the file-drawer
and provide some guidance for future research.
\end{abstract}
{\bfseries \emph Keywords}
\def\sep{\textbullet\ }
List experiments \sep Sensitivity bias \sep Social desirability
bias \sep 
Simulation

\newpage

\section{Introduction}\label{introduction}

Sensitivity bias\footnote{This is also referred to as social
  desirability bias (Krumpal 2013; Fisher 1993). We use the broader term
  introduced in Blair, Coppock and Moor (2020) which encompasses
  misreporting for a variety of reasons not limited to social
  desirability.} has been recognized as a fundamental barrier in survey
research on a variety of phenomenon, spanning drug use, racial
prejudice, sexual preferences and voting behavior (Fisher 1993). The
problem is that since people do not want to openly admit to sensitive
traits --opinions and behaviors considered socially undesirable,
offensive, or otherwise risky to reveal-- estimates of the population
prevalence of these traits will be biased.

The list experiment is a popular technique for mitigating sensitivity
bias. A list experiment provides survey respondents with additional
confidentiality by embedding the sensitive question within a list of
non-sensitive items, making it impossible to identify individual
responses to the sensitive question. Assuming that people answer
truthfully under the guise of privacy, list experiments will produce
unbiased estimates of sensitive traits. While a wealth of methodological
work has refined the design and analysis of list experiments, we call
attention to an important complication: the polarity (sign) of
sensitivity bias can vary across respondent subgroups, even cutting in
opposite directions.

For example, consider a survey question measuring belief in claims of
fraud in the 2020 US Presidential Election. Belief in Trump's claims of
fraud may be a sensitive trait for Democrats, who are likely reluctant
to admit they believe Democratic President-elect Joe Biden's win was
illegitimate. On the other hand, Republicans may not interpret belief in
electoral fraud as a sensitive topic. Alternatively, it could be the
case that Republicans are reluctant to admit that they \emph{do not}
believe Trump's claims.\footnote{For example:
  \url{https://nytimes.com/2023/02/16/business/media/fox-dominion-lawsuit.html}}
In either case, the sign of sensitivity bias would differ across
Democrats and Republicans. The bias from sensitivity for Democrats would
be negative, while for Republicans it would be zero or positive. We
refer to the situation where the (expected or observed) sign of
sensitivity bias differs across subgroups as non-uniform polarity.

The idea of non-uniform polarity is not entirely new. Methodologists
have pointed out that the possibility exists (e.g., Aronow et al. 2015,
58; Blair, Coppock, and Moor 2020, 1307; Chou 2018). There is also some
empirical support for the phenomenon in prior research: subgroup
analyses in list experiments often show apparent differences in
sensitivity bias across groups defined by characteristics such as
partisanship, education, gender (e.g., Blair and Imai 2012; Heerwig and
McCabe 2009; Kim and Kim 2016; Carkoglu and Aytaç 2015; Gonzalez
Ocantos, Jonge, and Nickerson 2014; Hatz, Fjelde, and Randahl 2023).
While the differences in these studies are not large enough to show with
statistical conventional statistical certainty that the sensitivity bias
cuts in opposite directions across subgroups, they do provide an
indication that non-uniform polarity may indeed exist as an empirical
phenomenon. Additionally, when violations of the monotonicity assumption
are detected in list experiments this may lead the researchers to
falsely conclude that the list experiment has failed, when in fact the
problem is non-uniform polarity, leading to the researchers not
publishing the results. The absence of statistically significant
empirical evidence of non-uniform polarity in the literature may be thus
be due to a `file-drawer' problem where studies which find non-uniform
polarity are misdiagonsed as failed list experiments and not published.

Drawing on the literature on social norms and preferences, there are
also good theoretical reasons to expect that group-based norms will
determine what is sensitive and what is socially desirable (e.g., Young
2015; Kuran 1990). Considering a range of sensitive topics-- e.g., vote
choice, prejudice, abortion and same-sex marriage-- it seems unlikely
that different social groups agree on what is socially acceptable and
what is sensitive. This may be especially true for topics where small
subgroups hold a belief that is contrary to the majority view and would
thus be considered sensitive. For instance, while we may theorize that
for a majority of respondents it would be considered sensitive to
express skepticism about the efficacy of vaccines, respondents who are
members of the broader anti-vaxxer community may consider it sensitive
to express a belief in the efficacy of vaccines. Similarly, for a
majority of respondents it would be considered sensitive to express a
belief that the world is flat, while for respondents who are members of
the broader flat earth community it may be considered sensitive to
express a belief that the world is round. In both cases, the sign of
sensitivity bias would differ across subgroups.

Despite the fact that the problem of non-uniform polarity has been
discussed in several research papers, to our knowledge, no systematic
research has been conducted to establish the consequences of non-uniform
polarity for a range of estimation goals and techniques. \emph{How do
the estimators we rely on for sensitive topics hold up when sensitivity
bias varies across subgroups? How large does a subgroup need to be to
shift estimates? Are some estimators more robust than others?}

This paper calls attention to the issue and explores its consequences.
We first discuss the potential problems that can arise from non-uniform
polarity. We describe how essential list experiment assumptions, or
tests of their violation, rely on an underlying implicit assumption of
uniform polarity. This implies that non-uniform polarity can cause
commonly used diagnostic tests to falsely detect violations of list
experiment assumptions. Second, estimation techniques relying on these
assumptions are likely to yield biased estimates. This concerns maximum
likelihood (ML) model estimates of the prevalence of a sensitive trait,
as well as the correlates of a sensitive trait. Third, estimates of
sensitivity bias itself will be biased in the presence of non-uniform
polarity. For example, positive and negative biases on the subgroup
level could cancel each other out, leading to an aggregate estimate of
no sensitivity bias.

To study the consequences of non-uniform polarity we set up a simple
Monte Carlo simulation. The simulation assesses the reliability of a
general diagnostic test, as well as bias in three commonly estimated
quantities of interest: the prevalence of the sensitive trait, the
correlates of the sensitive trait, and the extent of sensitivity bias.
We find that when the polarity of sensitivity bias is the opposite
across subgroups, Aronow et al's (2015) joint placebo test falsely
indicates the violation of list experiment assumptions. We also show
that, in the presence of non-uniform polarity: ML estimators combining
direct- and list experiment-measures generally return positively biased
estimates of the prevalence of a sensitive trait, the estimated
coefficients of covariates are biased in standard ML and combined ML
models under certain conditions, and estimates of sensitivity bias are
biased towards zero. For all three quantities, the bias is worse when
polarity runs in opposite directions across subgroups, and as the
difference in subgroup sizes increases.

From our simulation results, we provide some suggestions for researchers
when designing and analyzing list experiments. First, given the
limitations of existing diagnostic tests in detecting non-uniform
polarity, we urge researchers to theorize about potential subgroup
variation in sensitivity bias and to conduct confirmatory subgroup
analyses. Second, researchers could design list experiments to minimize
bias from non-uniform polarity, for example by phrasing the sensitive
item so that it is non-sensitive to one subgroup. Third, it's crucial to
understand that biases can emerge in both the prevalence estimate and in
covariate estimates, especially when covariate effects differ across
subgroups. This calls for a nuanced approach in theorizing about
sensitive traits and their correlates across different groups. Overall,
understanding the theoretical underpinnings of non-uniform polarity in a
given sample can help guide design and analysis choices.

Most broadly, the results have implications for political science and
survey research. Non-uniform polarity could explain why some list
experiments appear to `fail': diagnostic tests could return false
positives and countervailing biases across subgroups could net out,
leading researchers to incorrectly infer their experiment failed to
recover the true population prevalence. As others have pointed out
(Blair, Coppock, and Moor 2020; Gelman 2014; Hatz, Fjelde, and Randahl
2023), this can lead to a file-drawer problem, whereby list experiments
with unexpected results are not submitted for publication. This suggests
that we may know little about the prevalence of traits when sensitivity
bias moves in unexpected ways or is complex, perhaps varying along more
than one dimension.

This paper makes specific but important contributions. The list
experiment is widely used to measure public opinion on sensitive topics.
There has been a great deal of attention to the properties and
assumptions of list experiment estimators, and several important
advancements have been made. We focus on a concern that has been raised
in concluding paragraphs (Aronow et al. 2015; Hatz, Fjelde, and Randahl
2023) and on blogs (Gelman 2014), yet has not been investigated
systematically. Our aim in this paper is to call attention to the
problem and anticipate its consequences via simulation. We believe
further theoretical, empirical and methodological research is needed,
particularly employing real-world data, in order to better understand
the issue and develop solutions.

\section{The problem of non-uniform
polarity}\label{the-problem-of-non-uniform-polarity}

List experiments are a popular survey technique used to get accurate
estimates of sensitive traits: opinions and behaviors people don't want
to reveal openly. In a list experiment, the sensitive trait is embedded
in a list of non-sensitive traits and survey respondents are asked only
to report how many statements apply, but not which ones. This makes it
impossible to identify individual responses to the sensitive topic and
offers respondents privacy. By randomly assigning respondents to receive
the sensitive statement (treatment) or not (control), researchers can
calculate an aggregate prevalence rate using the difference-in-means
(DiM) across treatment and control. The logic of list experiments is
that respondents will answer truthfully when afforded privacy, allowing
researchers to avoid the bias from respondent misreporting.

List experiments thus produce unbiased estimates under the assumption
that respondents will tell the truth. However, as Ahlquist (2018) points
out, this is a fundamental contradiction, since list experiments are
used in a context where we expect respondents to lie about their beliefs
or experiences. Also, the list experiments offer only imperfect privacy:
answering affirmatively to all the statements would reveal the
respondent possesses the sensitive trait. More precisely, list
experiments rely on the assumption that respondents will only lie in
certain (strategic) ways.

Our focus in this paper is to point out that this assumption in turns
depends on the assumption that the polarity (sign) of sensitivity bias
is \emph{uni-directional} for a given topic and \emph{uniform} across
subgroups. That is, we must assume either that possessing the trait
(i.e.~answering affirmatively on the sensitive item) is the sensitive
response, or that not possessing the trait (i.e.~answering negatively on
the sensitive item) is the sensitive response. This means we expect
sensitivity bias to be \emph{either} negative or positive: polarity is
uni-directional for the survey topic (Hatz, Fjelde, and Randahl 2023;
Gelman 2014). We also must assume that the same response is sensitive
across subgroups defined by partisanship, gender, and so on. This means
we expect sensitivity bias will be of the same sign (positive or
negative) for all subgroups: polarity is uniform. With these assumptions
in place, we know what types of (strategic) lying to expect, and how to
test for unwanted types of lying (nonstrategic lying/measurement
error).\footnote{We think of uni-directional polarity as an assumption
  at the topic-level. For a given topic Z, researchers usually
  articulate uni-directional expectations: Z is a sensitive trait,
  therefore answering affirmatively to Z is the sensitive response, and
  we expected Z to be under-estimated in the aggregate (Blair, Coppock,
  and Moor 2020; Gelman 2014; Ahlquist 2018; Hatz, Fjelde, and Randahl
  2023). We think of uniform polarity as an assumption at the
  group-level. For topic Z and groups A and B, researchers usually
  articulate uniform expectations: Z is a sensitive trait to groups A
  and B, therefore answering Z=1 is the sensitive response for A and B,
  and we expected Z=1 to be under-estimated in the aggregate. The
  assumption that sensitivity bias is uniform thus also implies the
  assumption that it is uni-directional.}

To illustrate the essential role of these underlying polarity
assumptions, consider the essential \emph{no liars} assumption: this
states that respondents will not lie about the sensitive item in the
list experiment (Blair and Imai 2012, 51--52). However, given that
privacy fails if respondents answer yes to all statements, respondents
may in fact lie strategically by avoiding the maximum value of
statements. One way to test whether the \emph{no liars} assumption holds
is by looking for `ceiling' liars in the sample: respondents who choose
one less than maximum value in the list experiment (Blair and Imai 2012;
Ahlquist 2018). However, as Hatz et al (2023) point out, what we
consider a `floor' or `ceiling' liar depends on what we as researchers
assume is the sensitive response (Hatz, Fjelde, and Randahl 2023,
Appendix B.4). And, if we are wrong about which response is sensitive,
or if which answer is sensitive varies across subgroups, then the tests
for `floor' and `ceiling' liars may falsely detect violations of the
no-liars assumption.

Similar issues arise when looking for `top' and `bottom' coder error
processes in the data, i.e.~respondents who simply select the maximum or
minimum value on the list experiment. Ahlquist (2018) notes that `top'
and `bottom' error processes may be either strategic or non-strategic.
The error process is non-strategic if the respondent simply chooses the
maximum or minimum value out of inattentiveness, but it may be strategic
if the respondent chooses the maximum or minimum value in order to avoid
revealing the sensitive trait. Following the same logic as tests for
floor- and ceiling liars, whether we interpret the presence of `top' or
`bottom' lying as strategic or as a violation of assumptions depends on
our assumptions about which response is the sensitive response. In the
presence of non-uniform polarity, this test could misdiagnose the type
of error process.

Extending the same logic, consider the \emph{monotonocity} assumption
introduced by Aronow et al (2015). The monotonicity assumption states
that individuals will not \emph{falsely confess} to a sensitive trait in
a \emph{direct} measure. This a fairly straightforward assumption of
rationality: why falsely confess, in a setting without additional
privacy, to an opinion that is socially undesirable? However, whether
this assumption is violated or not again depends on our polarity
assumptions. In the presence of non-uniform polarity where \emph{not}
possessing the trait is the sensitive response for one group, then
\emph{false confessions} would be rational (strategic lying) for this
group \footnote{Hatz, Fjelde, and Randahl (2023) elaborate this point in
  Appendix B.4}. Again, if we are incorrect about what is sensitive, or
if what is sensitive varies across subgroups, then the diagnostic test
could indicate non-strategic measurement error while at least for some
respondents the misreporting is strategic. Additionally, considering a
subgroup for whom \emph{not} possessing the trait is the sensitive
response, the monotonicity violation might instead be termed a
\emph{false denial}: it would be irrational to falsely deny possessing
the trait. Under the uni-directional and uniform polarity assumption
that possessing the trait is sensitive, however, we would not detect
this form of non-strategic measurement error while testing the
monotonicity assumption.

Another way to think about the problem is that list experiments rely on
an underlying assumption of `one-sided lying': we assume lying will only
occur in the direction specified by our theoretical assumptions about
the sensitive topic (Ahlquist 2018). If we observe lying in the opposite
direction (e.g., floor effects when we expect ceiling effects, or false
confessions) then we interpret these as non-strategic error processes,
which violate list experiment assumptions about specific forms of
(strategic) lying.

The additional point we are trying to emphasize is that since we assume
one-sided lying for the entire sample we apply diagnostic tests on the
sample-level, so if one relatively large subgroup lies in the `wrong'
direction then the diagnostic test may incorrectly suggest the list
experiment as whole fails to meet design assumptions. The implication is
thus that non-uniform polarity can cause commonly used diagnostic tests
to falsely detect violations of essential list experiment assumptions.
Here, when we say falsely detect, we mean that the test may mislabel
response behavior that is rational for some respondents as
non-strategic. A diagnostic test could also fail to detect true
violations of assumptions among subgroups, via the same mechanism of
testing at the aggregate level and misdiagnosing the type of error at
least for some respondents.

In addition to being problematic for diagnostic tests, the more general
problem is that list experiments rely on assumptions such as no liars
and monotonicity in order to produce unbiased estimates. This means that
when these assumptions are violated (as they appear to in the presence
of non-uniform polarity) then list experiment estimators which rely on
these assumptions will be biased. In particular, this has implications
for maximum likelihood (ML) estimators which rely on the monotonicity
assumption, such as estimators which combine the list experiment and
direct measure into a combined measure of prevalence Eady
(2017).\footnote{The monotonicity assumption must be met in order to
  model responses to the list experiment and direct measure
  simultaneously, such as when calculating a combined measure of
  prevalence. When this assumption holds, we can treat reports of
  sensitive opinions in the direct measure as the truth. This makes it
  possible to combine list experiment and direct measures to get a
  single prevalence estimate (Aronow et al. 2015). It also must be met
  in order to model the effect of covariates on the likelihood of the
  respondent concealing the sensitive trait in the direct measure, as in
  Eady's (2017) misreporting estimator.} In fact, combined ML estimators
rely on the monotonicity assumption to the extent that `false
confessors' must be identified and removed from analysis prior to
estimation.\footnote{False confessors can be identified by looking for
  `top-coders': respondents who answer the maximum number of items in
  the list experiment (i.e., confess) yet answer negatively to sensitive
  item in the direct measure (i.e., they falsely confess). However, as
  we have explained, this diagnostic requires the assumption that
  possessing the trait is the sensitive response (i.e, assuming polarity
  is uni-directional and uniform).}

In specific situations, non-uniform polarity may also have implications
for the standard ML estimator (Blair and Imai 2012), which relies only
on the no liars assumption. In particular, if polarity varies across
subgroups, and a covariate which predicts the (assumed) sensitive trait
has \emph{different effects} across subgroups, then this may result in
biased estimates of the covariate. This is the case because the model
would be unable to separate the effects between the two (or more)
groups. For example, consider the covariate ``political engagement'' in
the initial example of measuring belief in Trump's claims of electoral
fraud. Assume that among Democrats, political engagement is negatively
correlated with belief in Trump's claims and assume that among
Republicans, political engagement is positively correlated with belief
in Trumps claims. Since the covariate is essentially co-linear with the
variable identifying the respondent subgroups, the model will produce
biased estimates of the effect of the covariate. While this is in
essence a problem of model misspecification it is nonetheless not a
trivial one, as solving this would both require knowing that polarity
varies across subgroups, and having enough respondents to ensure that
the estimation is still properly powered.

It follows from this discussion that estimates of sensitivity bias
itself are likely to be biased. Estimating sensitivity bias may be of
theoretical interest (for example, see Lax and Stollwerk 2016; Blair,
Coppock, and Moor 2020) or used to validate whether the list experiment
successfully reduced misreporting (e.g., Kramon and Weghorst 2019).
Sensitivity bias is commonly operationalised as the difference between
the list experiment estimate and the direct question estimate (Blair,
Coppock, and Moor 2020). However, if list experiment assumptions do not
hold, and the estimated prevalence is biased, then so will be the
estimate of sensitivity bias. Further, in the particular case of
non-uniform polarity there are intuitive reasons estimates of
sensitivity bias could be off, as others have speculated (e.g., Lax and
Stollwerk 2016; Hatz, Fjelde, and Randahl 2023). For example, if
sensitivity bias is positive within a large respondent subgroup, then
the overall estimated sensitivity bias will be positive although
sensitivity bias may negative or null for other subgroups. Or, if
sensitivity bias cuts in opposite directions across subgroups, the bias
may net-out. In this case, a topic which is highly sensitive for each
subgroup may appear non-sensitive in the aggregate.

\section{Simulation study}\label{simulation-study}

To study the consequences of non-uniform polarity, we conduct a small
Monte Carlo simulation experiment.

Our simulation is set up as follows. We are interested in the sensitive
topic \(Z\) and we assume possessing the trait (\(Z=1\)) is sensitive.
We generate data from a standard list experiment with 2000 respondents,
of which 1000 are in the list experiment treatment group and 1000 are in
the list experiment control group. There are two subgroups in the
experiment, group A and group B. In group A, possessing the trait
(\(Z=1\)) is the sensitive response, while in group B not possessing the
trait (\(Z=0\)) is the sensitive response. Given this non-uniform
polarity, the monotonocity assumption may appear violated in the
aggregate, since some respondents will appear to \emph{falsely confess}
in the direct measure. However, we ensure monotonicity holds on the
subgroup level, i.e.~group A will not falsely confess to \(Z=1\) in the
direct measure and group B will not falsely confess to \(Z=0\). We also
ensure that the following list experiment assumptions hold: no liars, no
design effects and treatment independence. Apart from group membership
(\(x_1\)), we include two covariates, \(x_2\) and \(x_3\), where \(x_2\)
predicts the likelihood of the respondent having the trait \(Z=1\) and
\(x_3\) affects the likelihood of respondents lying in the direct
measure. The true proportion of respondents with \(Z=1\) is fixed to
25\% in both group A and group B.\footnote{This may be considered
  unreasonable as then the vast majority of respondents in group B would
  hold the, to them, undesirable trait, but allows for a better
  illustration of the effect of the problem.}

In the simulations we vary several conditions. First, we vary the type
of non-uniform polarity. In one set of simulations, \(Z=1\) is the
sensitive response to group A and \(Z=0\) is the sensitive response to
group B, so that sensitivity bias is positive for group A and negative
for group B (i.e.~the polarity of sensitivity bias is the opposite for
group B). In the other set of simulations, \(Z=1\) is the sensitive
response to group A and \(Z\) is non-sensitive to group B, so that
sensitivity bias is positive for group A and zero for group B (i.e.,
there is no sensitivity bias in group B).

Second, we vary the proportion of respondents in group B from 0 to 50\%.

Third, we vary whether the effect of covariate \(x_2\) on \(Z\) is the
same or different across the subgroups. In one set of the simulations,
\(x_2\) has the same effect on the likelihood that respondents have
\(Z=1\) in both group A and group B (i.e.~the correlation between
\(x_2\) and \(Z\) goes in the \emph{same direction} for group A and B).
In the other set of simulations, the covariate \(x_2\) predicts \(Z=1\)
for group A and \(Z=0\) for group B (i.e.~the correlation between
\(x_2\) and \(Z\) goes in \emph{the opposite direction} for group A and
B).

In each simulation, we estimate the p-value for Aronow et al's (2015)
joint placebo test for the no liars, no design effect, treatment
independence and monotonicity assumptions. Since the simulations are set
up to ensure no design effects, treatment independence and monotonicity
on the sub-group level, we can interpret this as a test of the
monotonicity assumption on the aggregate level, under the underlying
polarity assumption that \(Z=1\) is the sensitive response for all. We
then estimate the bias in three common quantities of interest: 1) the
estimated prevalence of \(Z=1\) using the difference in means (DiM)
estimator, the standard maximum likelihood (ML) estimator (Blair and
Imai 2012) and the estimator combining the direct and list experiment
measures (Aronow et al. 2015; Eady 2017) (Combined ML\footnote{In order
  to estimate the combined ML estimator in the presence of apparent
  `false confessors', as implied by non-uniform polarity, we run the
  combined ML model while including one extra control item which is
  always 0. This ensures total anonymity for top-coders, so they cannot
  appear as `false confessors'. The extra item does not affect the bias
  calculations in any way. This change only affects the variance of the
  estimates, which are not being analyzed in the study.}); 2) the beta
coefficients for the effect of covariates \(x_1\), \(x_2\), and \(x_3\)
on the likelihood of \(Z=1\) from the maximum likelihood methods; and 3)
the aggregate sensitivity bias (the estimated prevalence of \(Z=1\)
using DiM minus the estimated prevalence in the direct measure).
Although we do not expect the DiM estimator or the ML estimator to
produce biased estimates for the proportion of respondents with \(Z=1\)
as they do not rely on the monotonicity assumption, we include these for
reference.

For each setup we run the simulation 1000 times and report the p-value
for the diagnostic test as well as the bias introduced in the different
quantities of interest.

\section{Results}\label{results}

\subsection{Aronow et al's joint placebo
test}\label{aronow-et-als-joint-placebo-test}

Figure 1 shows the p-values for Aronow et al's (2015) joint placebo test
for the no liars, no design effect, treatment independence and
monotonicity assumptions. We see that when the polarity of sensitivity
bias is the opposite sign for group B, the test rejects the null
hypothesis that all four assumptions hold. From our simulation set-up,
when know that the test is flagging a violation of monotoncity on the
aggregate level, although we know monocity holds on the subgroup level.
This result is consistent once group B reaches around 25\% of the
sample. The test does not reject the null when \(Z\) is not sensitive to
group B beyond what is expected by random chance.

\begin{figure}[H]

{\centering \includegraphics[width=0.6\textwidth,height=\textheight]{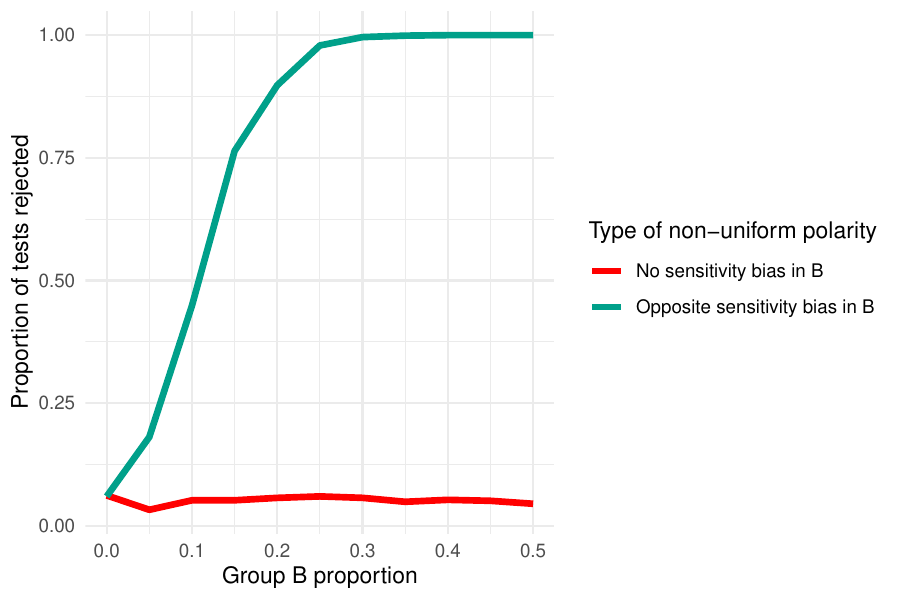}

}

\caption{P-values for Aronow et al's (2015) joint placebo test}

\end{figure}%

\subsection{Estimated prevalence of the sensitive
trait}\label{estimated-prevalence-of-the-sensitive-trait}

Figure 2 shows the bias in the estimated proportion of respondents with
\(Z=1\). From this figure it is clear that since there is no violation
of the no liars assumption, both the DiM and the standard ML estimators
give unbiased estimates of the true proportion with \(Z=1\) across all
specifications. However, since the monotonicity assumption is violated
in the aggregate, the ML model which combines the direct and list
experiment measures generates positively biased estimates of the
prevalence of \(Z=1\). There is an exception in the special case where
\(Z\) is not sensitive to group B \emph{and} the effect of \(x_2\) is
the same across both subgroups.

\begin{figure}[H]

{\centering \includegraphics[width=0.8\textwidth,height=\textheight]{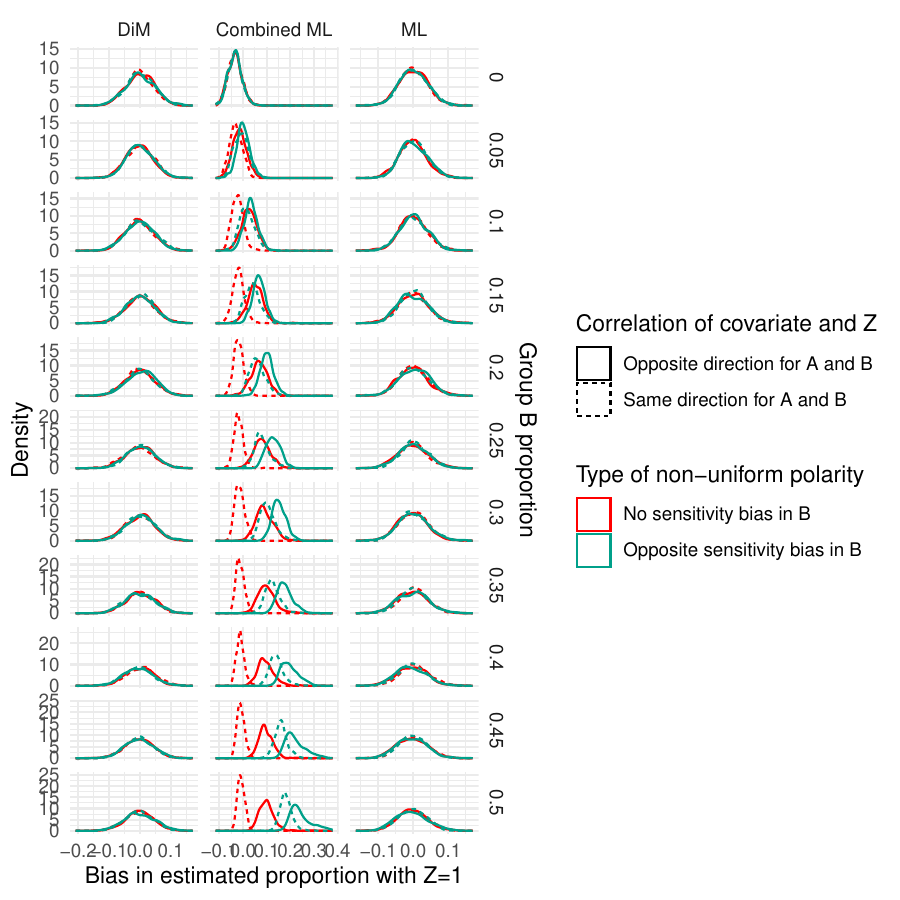}

}

\caption{Bias in the estimated prevalence of the sensitive trait}

\end{figure}%

\newpage{}

\subsection{Estimated coefficients of
covariates}\label{estimated-coefficients-of-covariates}

Figure 3 shows the bias in the estimated beta coefficient for the effect
of covariate \(x_2\) on the likelihood of the respondent having \(Z=1\).
Here we can see that both the standard ML model and the combined ML
model produce biased beta coefficient estimates for \(x_2\). There is an
exception for the standard ML model when the effect of \(x_2\) is
identical across both subgroups and for the combined ML model when the
topic is not sensitive in group B \emph{and} the effect of \(x_2\) is
the same across groups.

\begin{figure}[H]

{\centering \includegraphics[width=0.8\textwidth,height=\textheight]{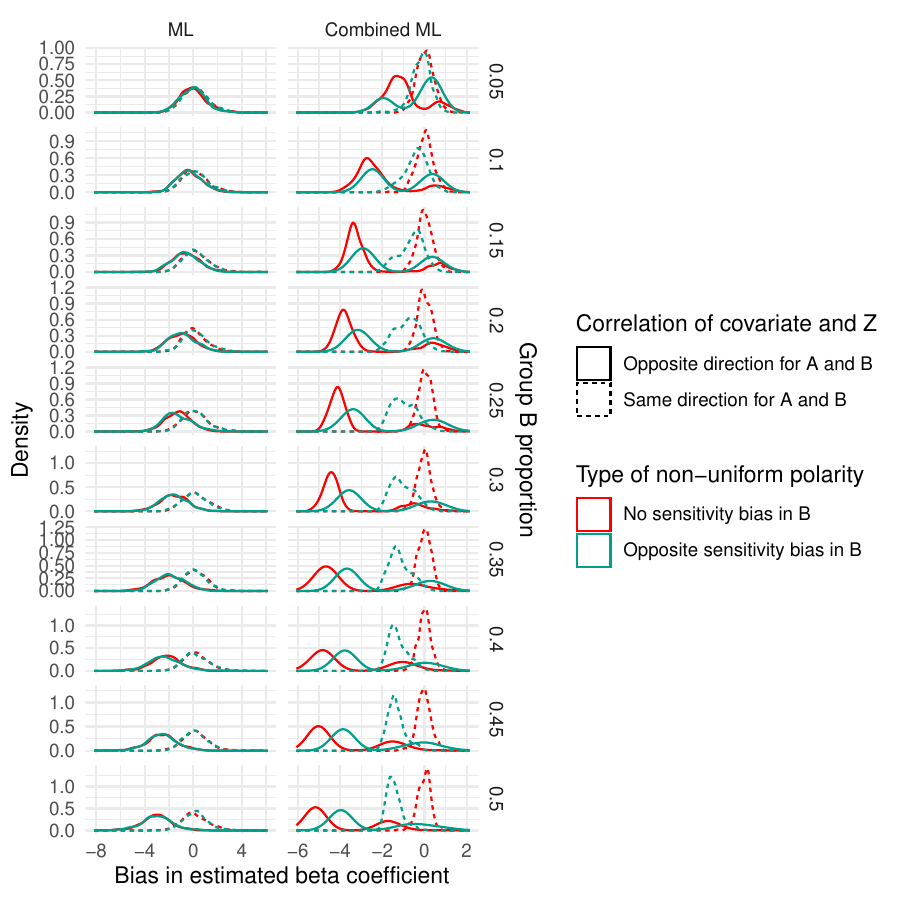}

}

\caption{Bias in the estimated beta coefficient for the effect of x2 on
the likelihood of the sensitive trait}

\end{figure}%

\newpage{}

\subsection{Sensitivity bias}\label{sensitivity-bias}

\begin{figure}[H]

{\centering \includegraphics[width=0.8\textwidth,height=\textheight]{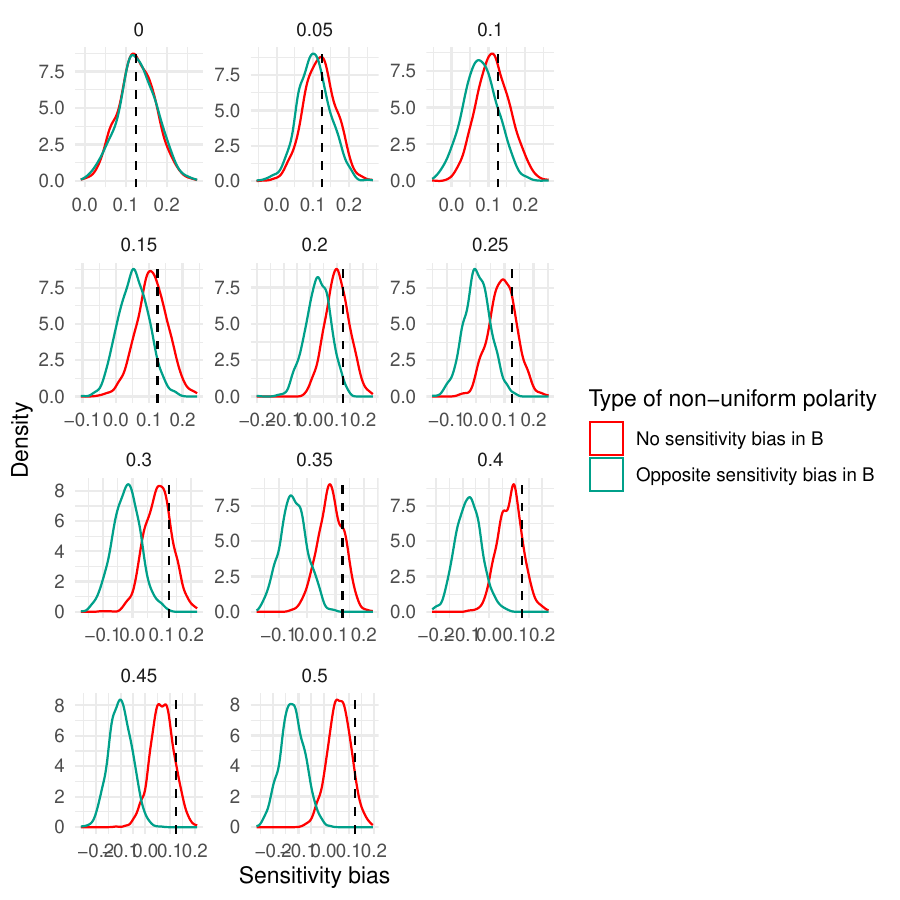}

}

\caption{Bias in the estimated sensitivity bias}

\end{figure}%

Figure 4 shows how estimates of sensitivity bias (the estimated
prevalence of \(Z=1\) using DiM minus the estimated prevalence in the
direct measure) are affected by non-uniform polarity. We see that even
when the size of group B is relatively small, the estimated sensitivity
bias shrinks quickly towards zero. The problem is much more severe when
the polarity is of opposite signs across group A and group B, but is
also noticeable when \(Z\) is not sensitive to group B.

\section{Discussion}\label{discussion}

The results from the simulation study largely confirm our expectations.
Under the simulated condition of subgroups with opposite polarities,
Aronow et al's (2015) joint placebo test detects a violation of the
monotonicity assumption on the aggregate level, although the simulations
ensure monotonicity holds on the sub-group level. This result is
consistent once the size of the subgroup becomes substantial (at 25\% of
the sample). However, the test does not flag a monotonocity violation in
the simulated case that the topic is not sensitive to one group.

We also find that list experiment estimators return biased estimates of
three common quantities of interest. ML estimators combining direct and
LE-measures generally return positively biased estimates of the
prevalence of a sensitive trait, the estimated coefficients of
covariates are generally biased in standard ML and combined ML models,
and estimates of sensitivity bias are attenuated. For all three
quantities, the bias is worse when polarity runs in opposite directions
across subgroups, and as the difference in group sizes increases.

The headlines are perhaps not surprising. Since having a subgroup with
the opposite polarity patently violates the monotonicity assumption, a
general diagnostic test will point to list experiment failure. It is
also not surprising that combined ML estimators, which rely on the
monotonocity assumption, return biased estimates in the presence of
non-uniform polarity: Aronow et al (2015) call attention to this problem
already in their initial paper. And as many intuit, the estimated
aggregate sensitivity bias shrinks when sensitivity bias takes opposite
signs across subgroups. Yet, we believe that these results highlight a
number of important points to keep in mind when designing, analyzing and
interpreting list experiments.

First, there is currently no diagnostic test for non-uniform
polarity.\footnote{Although it is possible to estimate sensitivity bias
  by subgroup in order to observe non-uniform polarity, issues of
  greater variability in list experiment measures and insufficient
  statistical power make this unreliable (Carkoglu and Aytaç 2015, 558;
  Blair, Coppock, and Moor 2020).} Aronow et al's (2015) placebo test is
a joint test of the monotonicity, no liars, no design effects, and
treatment independence. In a non-simulated setting, the researcher will
not be able to know which of the assumptions have been violated when the
test is rejected. \footnote{An additional problem is that the test has
  limited power, and will only reject the null when the sample size is
  sufficiently large.} Even if we could isolate the monotocity
assumption, as in our controlled simulations, the test will only flag
non-uniform polarity in the case of opposite polarities; it does not
detect non-uniform polarity where the topic is sensitive in one subgroup
but not sensitive in another. In the absence of post-hoc tests for
non-uniform polarity, we urge researchers to draw on theory. For
instance, are there theoretical reasons to believe that certain
subgroups have different norms or beliefs regarding the topic? Although
list experiments are often motivated by uni-directional assumptions
(Blair, Coppock, and Moor 2020; Gelman 2014; Ahlquist 2018; Hatz,
Fjelde, and Randahl 2023), we suspect that few topics that are truly
sensitive also correspond to universally-held values. In particular, we
believe that when the joint test is rejected that researchers should
consider the possibility of non-uniform polarity as the culprit and
revisit the theoretical expectations among subgroups. If theoretically
motivated subgroups with potentially different sensitivity bias are
identified, researchers should proceed with subgroup analyses of these
groups and investigate the potential for non-uniform polarity by running
the diagnostic tests separately for each subgroup. Ideally, researchers
should also pre-register expectations about subgroup variation in
sensitivity bias in order to motivate specific confirmatory sub-group
analyses while avoiding multiple comparisons.

Second, we have shown that bias in ML estimators is worse when polarity
runs in opposite directions across subgroups, and as the difference in
group sizes increases. At the design stage, researchers should consider
both the nature of non-uniformity and the relative size of relevant
subgroups. Since bias tends to be worse in the case of opposite
polarities, researchers may consider phrasing the sensitive item in a
list experiment so that it is non-sensitive to one subgroup. As even a
relatively small group with the opposite polarity can induce bias,
researchers should not discount minority groups.

Third, we've noted that bias occurs in estimates of prevalence and in
estimates of covariate effects when the effect of a covariate differs
across subgroups with different polarities. While this is essentially a
problem of model mis-specification rather than with the estimators
themselves, it is not unreasonable to believe that covariates which
predict the trait may differ across groups with different attitudes
toward the sensitive trait. Beyond theorizing around which groups may
differ in what they consider sensitive, it's important to think about
the correlates of the sensitive trait in each group.

Fourth, non-uniform polarity can cause researchers to draw incorrect
conclusions from the data. One reason is that diagnostic tests may
mislabel strategic response behavior among some respondents as
non-strategic. This may lead researchers to falsely conclude that the
list experiment has failed to satisfy key assumptions. It could also
lead to poor post-hoc analysis decisions. For example, removing `false
confessors' from the data may imply removing the rational respondents
while preserving inattentive respondents; removing strategic liars in
one subgroup while allowing non-strategic liars in another subgroup to
stay. In the presence of a large subgroup with the opposite polarity, it
could also mean removing a large share of respondents, reducing
statistical power and generalizability. Additionally, because diagnostic
tests rely on uni-directional and uniform polarity assumptions, and are
carried out on the sample-level, the tests cannot detect violations of
list experiment assumptions within respondent subgroups. Another reason
non-uniform polarity could cause researchers to draw incorrect
conclusions is that the estimated sensitivity bias may run in the
opposite direction as expected, or be close to zero. Since the presence
of sensitivity bias is often used to validate the list experiment as a
measurement technique, this could lead to the conclusion that the
experiment failed to encourage truthful responses. The longer-term
consequence of apparent list experiment failure is a `file-drawer
problem', whereby scholars with unexpected results chose not to pursue
publication (Blair, Coppock, and Moor 2020; Gelman 2014; Hatz, Fjelde,
and Randahl 2023).

Finally, as the non-uniform polarity only causes bias for the ML based
estimators we also urge researchers to always consider simpler
estimators, such as the difference-in-means estimator or standard ML
estimator, when analyzing list experiments in cases where there is a
theoretical reason to believe that the monotonicity assumption
\emph{could be} violated. These estimators do not rely on the
monotonicity assumption and are therefore unbiased in the presence of
non-uniform polarity. If there are large differences in the estimated
aggregate sensitivity bias across subgroups, this may be an indication
that the monotonicity assumption is violated, and the simpler estimators
should be used instead.

This initial paper is only a first step in the direction of
understanding the implications of non-uniform polarity. Further
theoretical, empirical and methodological research is needed. Drawing on
the literature on individual-level preferences and social norms could
deepen our understanding of the roots and nature of divergent attitudes
towards sensitive topics. Empirical research could establish the extent
of non-uniform polarity in real-world survey data. Methodological work
may potentially develop new diagnostic tests and robust estimators, as
well better guidance for the design of list experiments in the context
of non-uniform polarity.

\newpage{}

\section*{References}\label{references}
\addcontentsline{toc}{section}{References}

\phantomsection\label{refs}

\end{document}